# Quantum information technology with Sagnac interferometer: Interaction-free measurement, quantum key distribution and quantum secret sharing

Wellington Alves de Brito and Rubens Viana Ramos

wbrito@deti.ufc.br          rubens@deti.ufc.br

Department of Teleinformatic Engineering, Faculty of Teleinformatic Engineering, Federal University of Ceará, 60455-760, C.P. 6007, Fortaleza-Ce, Brazil

**Abstract**

The interferometry of single-photon pulses has been used to implement quantum technology systems, like quantum key distribution, interaction-free measurement and some other quantum communication protocols. In most of these implementations, Mach-Zehnder, Michelson and Fabry-Perot interferometers are the most used. In this work we present optical setups for interaction-free measurement, quantum key distribution and quantum secret sharing using the Sagnac interferometer. The proposed setups are described as well the quantum protocols using them are explained.

# 1. Introduction

Quantum information technology is the new engineering area responsible for the experimental realization of quantum communication protocols and quantum computational circuits. However, in despite of the potentialities of quantum information to provide new ways of communication and computation, to work with quantum data is a hard task. For quantum gates implementations, several different technologies have been tested being optical and photonic devices [1-5], quantum dots [6], superconducting devices [7,8], semiconductors [9,10] and nuclear magnetic resonance [11-13] the most important and promising. On the other hand, for quantum communication purposes, optical and photonic technology is, up to now, the only one. This happens because, among other reasons, light polarization is a qubit relatively easy to create, to process and to detect, a photon can be sent far way in an optical fiber and interferometry of single-photons is a powerful technique to observe quantum phenomena. In fact, most experimentally realized quantum key distributions (QKD) setups were implemented using light polarization and/or single-photon interferometry (with weak coherent states). Further, the interferometry of single-photons can also be used for interaction-free measurement, whose goal is to identify the presence of an object without any interaction with the same. Most of the realizations of quantum technology using interferometry of single-photons have used Mach-Zehnder, Michelson or Fabry-Perot interferometers. In this work we present optical setups for interaction-free measurement, QKD and quantum secret sharing using the Sagnac interferometer. The proposed setups and the quantum protocols for their use are explained. This work is outlined as follows: In Section 2, interaction free-measurement with Sagnac interferometer is discussed. In Section 3, QKD using Sagnac interferometer is presented. In Section 4, the setup for quantum secret sharing between five persons using Sagnac is shown. At last, the conclusions are presented in Section 5.

## 2. Interaction-free measurement using Sagnac intereferometer

The fascinating experiment of interaction-free measurement consists in to identify the presence of an object in a determined place without interacting in anyway with the object. The key property that allows such task to be realized is the wave-particle characteristic of single particles like photons. This wave-particle behavior is readily observed in single-photon interferometry, in fact, the first interaction-free experiment was proposed in [14] using single-photons in a Mach-Zehnder (MZ) interferometer, as shown in Fig. 1

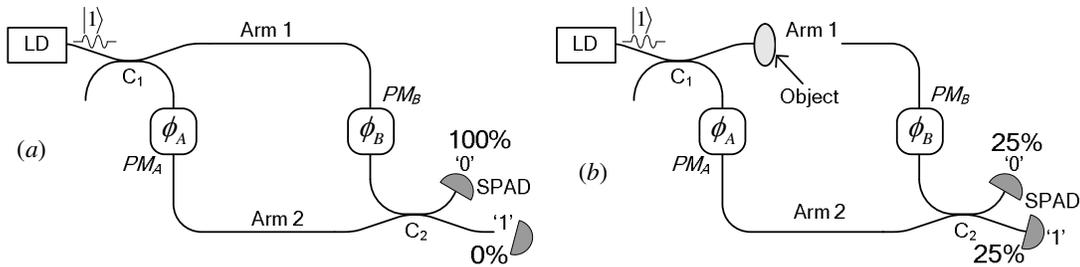

Figure 1 – Interaction free measurement using single-photon pulses and Mach-Zehnder interferometer having $\phi_A=\phi_B$. Part (*a*) – Object not inserted implies wave behavior. Part (*b*) – Object inserted implies particle behavior.

In Fig. 1 $C_1$ and $C_2$ are balanced optical couplers while SPADs are single-photon detectors. For the interaction-free experiment $\phi_A=\phi_B$, that is, when the object is absent the photon behaves like wave and it emerges always at the '0' output, as shown in part (*a*) of Fig. 1. On the other hand, when the absorber object is present in one of the interferometer arms, the photon will behave as a particle and it will be detected, with probability 25% at output '1', as shown in part (*b*) of Fig. 1. Hence, every time detection occurs at output '1', in an ideal noiseless system, one can be sure the object is present. The interaction-free experiment of Fig. 1 has low efficiency since the probability of getting a correct and conclusive result when the object is present, is only 25%. A higher performance interaction-free experiment using single-photon polarization was proposed in [15], and it can be seen in Fig. 2.

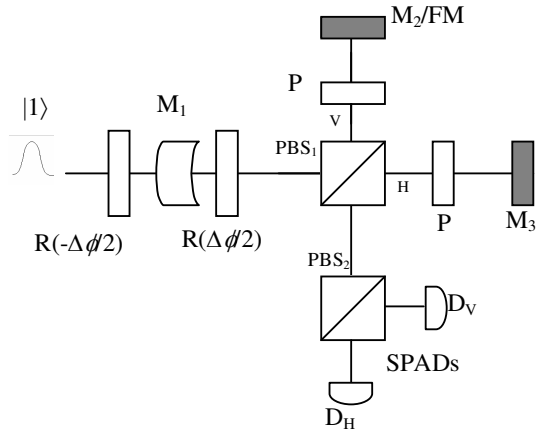

Figure 2: Interaction-free experiment using single-photon polarization. $PBS_{1(2)}$ – Polarizing beam splitter, $D_{H(V)}$ – SPDAs, $R(\pm\Delta\phi/2)$ – Polarization rotator, $M_1$ – Single-direction mirror, $M_{2(3)}$ – Mirror, FM – Faraday mirror, P – Pockels cell.

In Fig. 2, $R(\pm\Delta\phi/2)$ are polarization rotators of $\pm\Delta\phi/2$; $M_1$ is a single-direction mirror, that is, light is highly transmitted from left to right and highly reflected from right to left; $M_2$ and $M_3$ are common mirrors; FM is a Faraday mirror that rotates the input light polarization of $\pi/2$; P is a Pockels cell that rotates light polarization of $\pi/2$ when it is activated. The polarizing beam splitter guides the input light as shown in Fig. 3.

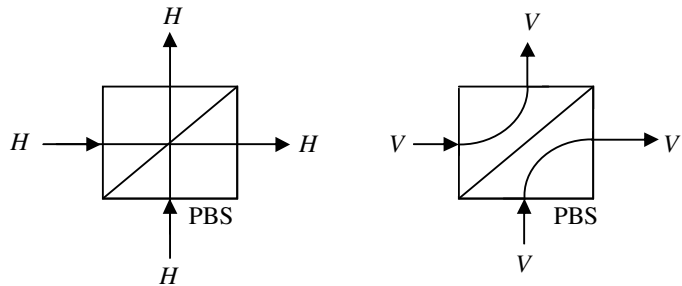

Figure 3: Guiding of input polarization states through 2x2 PBS.

The goal of the experiment in Fig. 2 is to determine, without interaction, which one is being used, $M_2$ or FM. When $M_2$ is being used the photon behaves like a wave and its polarization is rotated from horizontal to almost vertical due to successive actions of the polarization rotator $R(\Delta\phi/2)$. After $N$ runs, the photon polarization is $|N\Delta\phi\rangle$ that can be made very close to $|\pi/2\rangle$ (vertical), controlling $N$ and choosing $\Delta\phi$ properly. At this moment ($N\Delta\phi \sim \pi/2$) both Pocekls cells are activated. The photon arrives in $PBS_1$ coming from $M_2/M_3$ and it is guided to $PBS_2$ having polarization $|N\Delta\phi+\pi/2\rangle$ (almost horizontal) and it is detected in $D_H$ with probability very close to 100%. On the other hand, if FM is used the photon behaves like a particle. In this case, for each run, if the photon is not detected in $D_H$, one can be sure its polarization is $|0\rangle$, hence, the polarization rotation due to successive actions of $R(\Delta\phi/2)$ will not be accumulative and, after $N$ runs without detection in $D_H$, the polarization of the photon coming from $M_2/M_3$ will be horizontal. Activating the Pockels cells, the polarization will become vertical and it will be guided to $PBS_2$ and detected in $D_V$. Hence, in order to have a good performance for interaction-free measurement, the following conditions must be satisfied $N\Delta\phi\sim\pi/2$ and $[\cos^2(\Delta\phi)]^N\sim 1$, where this last condition is the probability of none detection in $D_H$ after $N$ runs when FM is being used. Choosing $\Delta\phi=\pi/(2N)$, $[\cos^2(\Delta\phi)]^N$ tends to $1-\pi^2/(4N)+O(N^{-2})$ for $N$ large. A third implementation of interaction-free measurement using Fabry-Perot (FP) interferometer was proposed in [16]. In this one, as shown in Fig. 4, the object is inserted or not inside the interferometer. If the object is absent, the FP interferometer has high transmissivity and the photons are detected in $D_1$. On the other hand, if the object is present, the FP transmissivity will be decrease and some photons will be detected in $D_2$.

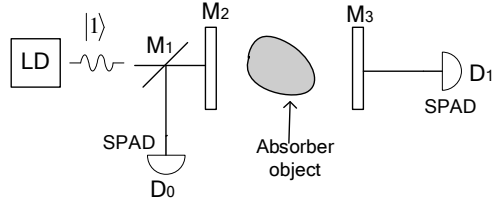

Figure 4: Interaction-free measurement using Fabry-Perot interferometer. $M_1$ – High reflectivity mirror.

Up to now one can see that interaction-free measurement experiments have been proposed using MZ, Michelson and FP interferometers. Now we show how to construct a interaction-free measurement experiment using the Sagnac interferometer [17]. The proposed setup can be seen in Fig. 5.

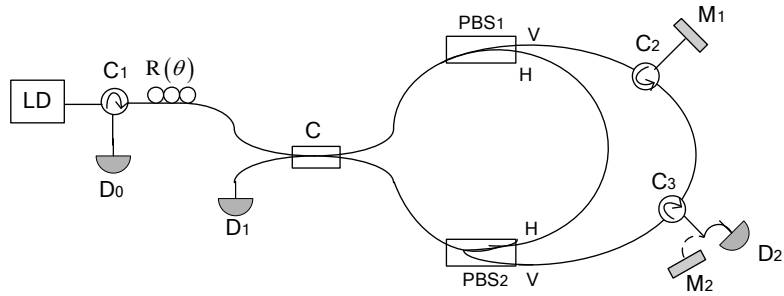

Figure 5: Interaction-free measurement using Sagnac interferometer. $R(\theta)$ is a polarization rotator and C is a balanced optical coupler.

The goal of the setup in Fig. 5 is to determine, without any interaction, which one is connected to circulator $C_3$, if mirror $M_2$ or detector $D_2$. For the correct functioning, the paths H-H and V-V must have the same length. The light emitted by diode laser LD is assumed to be horizontally polarized. Let us initially suppose that $M_2$ is connected. In this case, for any value of $\theta$, the photon will be behave like a wave and it will always be detected in $D_0$. On the other hand, if $D_2$ is connected, depending on $\theta$ value the photon will behave like wave ($\theta=0$), particle ($\theta=\pi/2$) or both at the same time ($0<\theta<\pi/2$). The probabilities of detection in $D_0$ ($P_0$), $D_1$ ($P_1$) and $D_2$ ($P_2$) are, respectively, given by

$$P_0 = \cos^2(\theta) + \frac{\operatorname{sen}^2(\theta)}{4} \tag{1}$$

$$P_1 = \frac{\operatorname{sen}^2(\theta)}{4} \tag{2}$$

$$P_2 = \frac{\operatorname{sen}^2(\theta)}{2} \tag{3}$$

Therefore, the probability of identifying that the device connected in $C_3$ is detector $D_2$ without losing the photon, that is, having detection in $D_1$, is given by (2), whose maximal value is 25% for $\theta=\pi/2$. Thus, the setup of Fig. 5 is, in terms of efficiency, equivalent to the setup shown in Fig. 1. However, using the setup presented in Fig. 6 one can determine the presence of $D_2$ without interaction with probability close to 1 for each photon used.

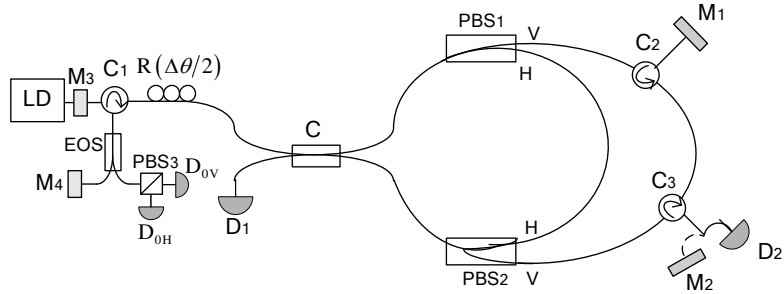

Figure 6: Efficient interaction-free measurement using Sagnac interferometer. $M_3$ single-direction mirror, EOS – electric-optical switch, $C_{1(2,3)}$ - circulators.

The functioning of setup in Fig. 6 is similar to the functioning of the setup presented in Fig. 2. Firstly, the photon emitted by laser source LD is horizontally polarized and the electro-optical key is connecting circulator $C_1$ to mirror $M_4$. If $M_2$ is connected, the photon comes into and leaves Sagnac interferometer several times and, for each time, its polarization is rotated of $\Delta\theta$ by polarization rotator $R(\Delta\theta/2)$. After $N$ runs, the photon polarization will be vertical or close, depending on the values of $N$ and $\Delta\theta$. At this moment ($N\Delta\phi \sim \pi/2$) the electro-optical key is switched connecting circulator $C_1$ to $PBS_3$. Thus, the photon will be guide by $C_1$ forward to $PBS_3$ and it will be detected in

$D_{0V}$ with high probability. On the other hand, if $D_2$ is connected, then photon polarization will not suffer accumulative rotation and, after $N$ runs, the electro-optical key is switched and the photon will be guided by $C_1$ to $PBS_3$ and detected in $D_{0H}$ with probability $\cos^2(\Delta\theta)$. As happen in the experiment of Fig. 2, the probability of the photon surviving after $N$ runs with $D_2$ connected is $[\cos^2(\Delta\theta)]^N$. There is an interesting difference in the performances of setups shown in Figs. 2 and 6. In Fig. 2, with FM connected, the probability of the photon to interact with FM per run is $\sin^2(\Delta\phi)$, while in Fig. 6, with $D_2$ connected, the probability of the photon to interact with $D_2$ is lower than $\sin^2(\Delta\theta)$. This happen because once the photon is vertically polarized, it can be clockwise or counter-clockwise. In this last case, the photon will be detected in $D_1$ with 50% of probability.

## 3. Quantum key distribution using Sagnac interferometer

Quantum key distribution (QKD) is the first quantum technology commercially available [18-20]. QKD experiments have been realized using MZ, Michelson and Sagnac interferometer. The first proposal of QKD using Sagnac, named circular type QKD, was proposed in [21]. The optical setup is shown in Fig. 7.

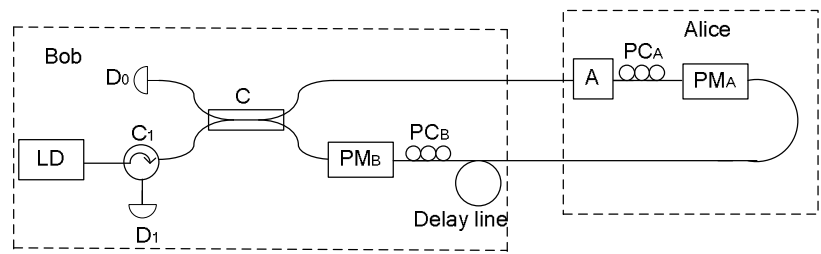

Figure 7: Circular type QKD. C – balanced optical coupler, PM – phase modulator, A – attenuator, PC – polarization controller.

The optical setup of Fig. 7 works as follows: Initially Bob sends a bright optical pulse. This pulse is split in two by the balanced optical coupler. One half going to Alice is clockwise ($P_{Clk}$) and the other half going to Alice is counter-clockwise ($P_{CClk}$). The pulse $P_{Clk}$ arrives first at Alice, since the pulso $P_{CClk}$ passes first by the delay line. Once in Alice, the pulse $P_{Clk}$ suffers attenuation in *A*, it has its polarization corrected by $PC_A$ and it passes by $PM_A$ without being modulated. Finally, it returns to Bob. Once in Bob, $P_{Clk}$ passes by the delay line, it has its polarization corrected by $PC_B$, it is phase modulated by $PM_B$ and, at last, arrives at optical coupler C. The pulse $P_{CClk}$ passes by $PM_B$ without being modulated, after it passes by $PC_B$, delay line and it follows to Alice. Once in Alice, $P_{CClk}$ is phase modulated by $PM_A$, it has its polarization corrected by $PC_A$, it is attenuated by *A* and it goes straight forward to optical coupler C at Bob. Both pulses arrive in C at the same and interference will take place. Depending on the phases difference applied by Alice in $P_{CClk}$ and by Bob in $P_{Clk}$, the photon will be guided to SPAD $D_0$ or $D_1$. Since both pulses take the same path, fluctuations of phase shifts are automatically compensated. The attenuation value of A is such that pulse $P_{CClk}$ leaves Alice having mean photon number close to 0.1. Another proposal of QKD using Sagnac interferometer was presented in [22]. In this one, an acoustic-optical phase modulator was used in Alice, making polarization controller easier, as well some care was taken in order to avoid a Trojan horse attack.

Differently of the setups proposed in [21,22] the setup proposed in this work is of the one-way type and, as happen with QKD using MZ interferometer, it is (ideally) naturally protected against Trojan horse attack. The setup of the proposed Sagnac-based QKD can be seen in Fig. 8. As can be observed, it uses light polarization and the Sagnac interferometer belongs only to Bob.

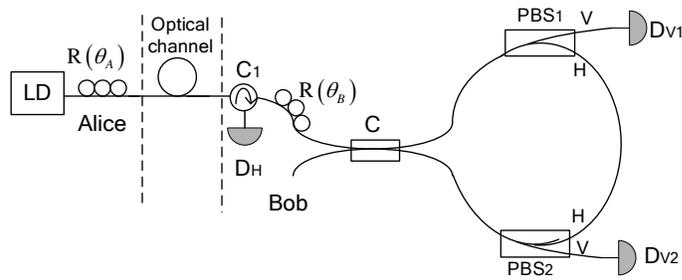

Figure 8: Optical scheme for polarimetric QKD using Sagnac interferometer.

The QKD protocol using setup of Fig. 8 works as follows: Alice sends single-photon pulses to Bob. For each pulse sent Alice choose randomly its polarization according to the codification: Basis 1 of Alice - {0 (0),π/2 (1)}; Basis 2 of Alice - {π/4 (0),3π/4(1)} (in X (Y), X is the polarization and Y the bit value it represents). For each photon that arrives at Bob, he applies a polarization rotation, randomly chosen, according to the codification: Basis 1 of Bob - {0 (0),-π/2(1)}, Basis 2 of Bob - {-π/4 (0),-3π/4(1)}. After transmission of all photons, Alice and Bob say publicly which bases they have used and, in the cases where Alice and Bob chose the same bases, Bob says to Alice in which detector he had detection $D_H$ or $D_{V(1\ or\ 2)}$. Having this information and knowing the polarization of her photon, Alice can discover which polarization, and hence the bit, Bob chose. In fact, when Alice and Bob choose the same bases ($\theta_A+\theta_B=0$ or $\pm\pi/2$) the photon impinging on the Sagnac has horizontal or vertical polarization. In the first case the photon behaves like wave suffering interference in C and being detected in $D_H$. In the second case, the photon behaves like particle and, if it is clockwise it will be detected in $D_{V1}$. If the photon is counter-clockwise, it will be detected in $D_{V2}$. When Alice and Bob choose wrong bases ($\theta_A+\theta_B=\pm\pi/4$) the photon behaves like wave and particle at the same time and it can be detected everywhere $D_H$, $D_{V1}$ or $D_{V2}$. As can be observed, this QKD protocol is close related to the wave-particle behavior. This does not happen with BB84 protocol in MZ interferometer, for example.

## 4. Secret sharing using Sagnac interferometer

At last, let us suppose the following problem: there exist a secret, a bit sequence $K$ of length $|K|$. This secret is shared among five persons in such way that none of them knows $K$. Each person has its own secret: Fred ($K_F$), Alice ($K_A$), Bob ($K_B$), Charlie ($K_C$) and David ($K_D$). The bits sequences obey the conditions $K \neq K_F \neq K_A \neq K_B \neq K_C \neq K_D$ and $|K|=|K_F|=|K_A|=|K_B|=|K_C|=|K_D|$. Fred is the one who will use the secret $K$, but he will need cooperation of his locally distant partners Alice, Bob, Charlie and David in order to obtain the correct secret $K$. This means that if one of the partners does not use its correct secret, Fred will, with high probability, not obtain the correct $K$. The optical setup of Fig. 9 can be used for such task.

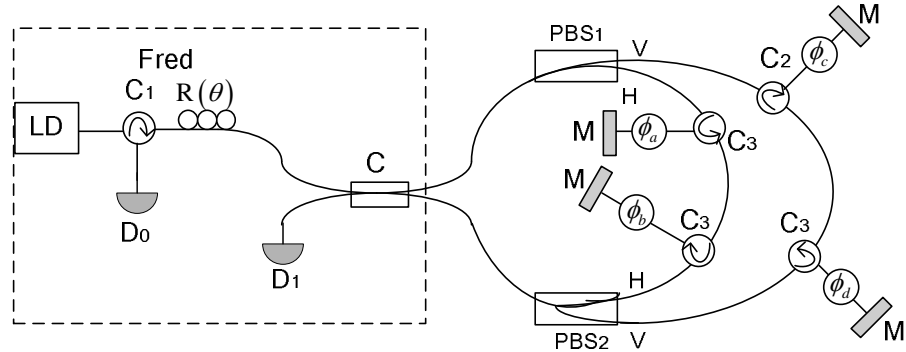

Figure 9: Optical setup using Sagnac interferometer for secret sharing between five persons. M – Mirror, R – polarization rotator, C – Optical coupler, $\phi_{a...d}$ – Phase modulators.

The probabilities of detection in $D_0$ ($P_0$) and $D_1$ ($P_1$) are given by:

$$P_0 = \cos^2(\theta)\cos^2\left(\frac{\phi_a - \phi_b}{2}\right) + \operatorname{sen}^2(\theta)\cos^2\left(\frac{\phi_c - \phi_d}{2}\right) \qquad (4)$$

$$P_1 = \cos^2(\theta)\operatorname{sen}^2\left(\frac{\phi_a - \phi_b}{2}\right) + \operatorname{sen}^2(\theta)\operatorname{sen}^2\left(\frac{\phi_c - \phi_d}{2}\right) \qquad (5)$$

When detection occurs in $D_0$ bit 0 is obtained while detection in $D_1$ implies bit 1. Observing (4) and (5), one can see that Fred choose who will define the bit value, Alice (*a*) and Bob (*b*), if Fred choose $\theta=0$ or Charles (*c*) and David (*d*), if Fred choose $\theta=\pi/2$. The possible values of the angles are $\theta \in \{0,\pi/2\}$, $\phi_{abcd} \in \{0,d\phi,2d\phi,3d\phi,...,Nd\phi,\pi, d\phi+\pi,2d\phi+\pi,3d\phi+\pi,...,Nd\phi+\pi\}$. The secrets that Alice, Bob, Charlie and David have are the sequences of phase shift that they have to apply. In order to have deterministic detections in Fred it is necessary to have $\phi_a$-$\phi_b$ and $\phi_c$-$\phi_d$ iqual to 0 or $\pi$ rad. Thus, the secrets that Alice and Bob have must be in such way that, if Alice has to apply the phase shift $kd\phi$, Bob's secret must indicate he has to use the phase shift $kd\phi$ or $kd\phi+\pi$, according to the bit value of the secret *K*, if 0 or 1, respectively. The same happens with Charlie and David. Hence, if, for example, David uses a different bit sequence, other than $K_D$, for those bits where Fred chose $\theta=\pi/2$, it may happen $\phi_c$-$\phi_d \neq 0$ and $\pi$. In this case, the photon will be detected in $D_0$ with probability $\cos^2[(\phi_c$-$\phi_d)/2]$ and it will be detected in $D_1$ with probability $\sin^2[(\phi_c$-$\phi_d)/2]$, meaning that an error can occur. If the real secret is a hash function of *K*, H(*K*), then even having few errors at the input, the output will be very different of the correct one.

## 5. Conclusions

We have discussed the use of Sagnac interferometer in quantum information technology. Three problems were discussed: interaction-free measurement, quantum key distribution and secret sharing. For the interaction-free measurement we present two optical setups, the first having 25% of success and the second almost 100% of success per photo used. Both are easily implemented using common linear optical devices. The QKD setup proposed is different from other proposals found in the literature since it is a one-way setup and, hence, it is more resistant against Trojan horse attack. Its disadvantage is the use of three SPAD and, since it uses single-photon polarization, it is suitable only for short distance and high transmission rate QKD in 850 nm. Further, the QKD protocol is a little different from BB84 since Bob has to inform to Alice the bases used and if detection occurred in $D_H$ or $D_V$ detectors (any of them). At last, the proposed QKD protocol is well related to wave-particle behaviour. Finally, we provided an optical setup for secret sharing between 5 persons. The secret, which is not known of any user, can be read or used by one of the partners, named Fred, only if all the other four partners collaborate using their correct individual secret. The proposed setup is easy to implement, being basically a Sagnac interferometer with polarization diversity.

## Acknowledgements

This work was supported by the Brazilian agency FUNCAP## References

[1] E. Knill, R. Laflamme and G. J. Milburn, 2001, *Nature*, 409, 46.

[2] T. B. Pitman, B. C. Jacobs and J. D. Franson, 2004, Los Alamos e-print quant-ph/0404059.

[3] F. M. Spedalieri, H. Lee, and J. P. Dowling, 2005, Los Alamos e-print quant-ph/0508113.

[4] T. C. Ralph, A. G. White, W. J. Munro, and G. J. Milburn, 2001, *Physical Review A*, 65, 012314.

[5] P. Kok, W. J. Munro, K. Nemoto, T. C. Ralph, J. P. Downling, and G. J. Milburn, 2005, Los Alamos e-print quant-ph/0512071.

[6] A. Imamoglu, 2003, *Physica E*, 16, 47.

[7] J. Q. You, J. S. Tsai, F. Nori, 2003, *Physica E*, 18, 35.

[8] Y. Makhlin, G. Schön, A. Shnirman, 2000, *Computer Physics Communication*, 127, 156.

[9] D. Copsey, M. Oskin, F. Impens, T. Metodiev, A. Cross, F. T. Chong, I. L. Chunag, and J. Kubiatowicz, 2003, IEEE Journal of Selected Topics in Quantum Electronics, 9, 6, 1552.

[10] S. Kawabata, 2004, *Science and Technology of Advanced Materials*, 295.

[11] I. L. Chuang, N. Gershenfeld, M. G. Kubinec, D. W. Leung, Proc. R. soc. Lond. A, 454, (1998) 447.

[12] I. L. Chuang, L. M. K. Vandersypen, X. Zhou, D. W. Leung, S. Lloyd, 1998, Nature, 393, 143.

[13] T. F. Havel, S. S. Somaroo, C.-H. Tseng, D. G. Cory, (2000) *Applicable Algebra in Engineering Communication and Computing*, 10, 339.

[14] A. C. Elitzur and L.Vaidman, 1993, *Foundations of Physics*, 23, 987.

[15] P Kwiat, H. Weinfurter, T. Herzog, A. Zeilinger, and M. A. Kasevich, 1995, *Physical Review Letters*, 74, 4763.

[16] T. Tsegaye, E. Goobar, A. Karlsson, G. Björk, M. Y. Loh, and K. H. Lim, 1998, *Physical Review A*, 57, 5, 3987.

[17] E. J. Post, Sagnac effect, 1967, *Review of Modern Physics*, 39, 2, 475.

[18] S. J. D. Phoenix and P. D. Townsend, 1995, *Contemporary Physics*, 36, 165.

[19] N. Gisin, G. Ribordy, W. Tittel and H. Zbinden, 2001, Los Alamos e-print quant-ph/0101098.

[20] M. Bourennane, D. Ljunggren, A. Karlsson, P. Jonsson, A. Hening and J. P. Ciscar, 2000, *Journal of Modern Optics*, 47, 2/3, 563.

[21] Tsuyoshi Nishioka, Hirokazu Ishizuka, Toshio Hasegawa, and Jun'ichi Abe, 2002, *IEEE Photonics Technology Letters*, 14, 4.

[22] Bing Qi, Lei-Lei Huang, Hoi-Kwong Lo, Li Qian, 2006, Los Alamos e-print quant-ph/0604187, 2006.